\documentclass[usenatbib,fleqn]{mnras}
\usepackage{amsfonts}
\usepackage{amsmath}
\usepackage{amssymb}

\usepackage[T1]{fontenc}

\usepackage[dvips]{graphicx}
\usepackage{myaasmacros}
\usepackage{color}
\usepackage{multirow}

\def\bea{\begin{eqnarray}}
\def\ena{\end{eqnarray}}

\title[How to make a mature accreting magnetar]
{How to make a mature accreting magnetar}
\author[A.P.~Igoshev \& S.~B.~Popov]{A.~P.~Igoshev$^{1}$\thanks{E-mail: ignotur@gmail.com}, S.~B.~Popov$^{2}$\thanks{E-mail: polar@sai.msu.ru}\\
$^{1}$Department of Astrophysics/IMAPP Radboud University Nijmegen, PO Box 9010, NL-6500GL Nijmegen, the Netherlands\\
$^{2}$Sternberg Astronomical Institute, Lomonosov Moscow State University, Universitetsky prospekt 13, 119234, Moscow, Russia\\
}

\begin{document}

\date{}

\pubyear{2017}

\maketitle

\label{firstpage}

\begin{abstract}
Several candidates for accreting magnetars have been proposed recently by different authors. Existence of such systems contradicts the standard magnetic field decay scenario where a large magnetic field of a neutron star reaches 
$\lesssim$~few~$\times 10^{13}$~G at ages $\gtrsim 1$~Myr. Among other sources, the high mass X-ray binary 4U0114+65 seems
to have a strong magnetic field around $10^{14}$ G. We develop a new Bayesian estimate for the kinematic age and demonstrate that 4U0114+65 has kinematic age 2.4-5 Myr ($95\%$ credential interval) since the formation of the neutron star. 
We discuss which conditions are necessary to
explain the potential existence of magnetars in accreting high-mass binaries with ages about few Myrs and larger. Three necessary ingredients are: the Hall attractor to prevent rapid decay of dipolar field, relatively rapid
cooling of the crust in order to avoid Ohmic decay due to phonons, and finally, low values of the parameter $Q$ to obtain long Ohmic time scale due to impurities.  
If age and magnetic field estimates for proposed accreting magnetars are correct, then these systems set the strongest limit on the crust impurity for a selected sample of neutron stars and provide evidence in favour of the Hall attractor. 
\end{abstract}


\section{Introduction}
\label{sec:intro}

 There are hundreds of known X-ray binaries with accreting 
neutron stars (NSs) in the Milky Way as well as in near-by galaxies
\citep{2006A&A...455.1165L, 2007A&A...469..807L, fabbiano2006, kundu}. 
In some cases it is possible to measure magnetic fields of compact objects
directly observing electron (or proton) cyclotron lines (see
\citealt{2015SSRv..191..293R} and references therein): $(E_\mathrm{cyc, e}/\mathrm{keV})=12(B/10^{12}\mathrm{G})\, (1+z)^{-1}$, where $B$ is the magnetic field, and $z$ is gravitational redshift at the line formation region. Typically, measured fields are in the range $10^{11}$~--~$10^{13}$~G (which is also determined by the energy range available for observational facilities, i.e. much smaller or larger fields correspond to lines out of the range of sensitivity of X-ray spectrometers $\sim 1-100$~keV). 
However, for majority of NSs the magnetic field can be estimated only with indirect methods based
on timing measurements (see Appendix B and, for example, \citealt{2015ApJ...813...91S, 2017MNRAS.470.3316D} and references therein). 
Among the latter cases there are a few
NSs for which estimates argue for magnetar scale fields
$\gtrsim 10^{14}$~G
\citep{2010A&A...515A..10D, 2012MNRAS.425..595R, 2012ApJ...757..171F,
2014MNRAS.437.3664H}. Such NSs have been called {\it accreting magnetars} (see Appendix A for a list of candidates).
Meanwhile alternative approaches, for example based
on a new model of settling wind accretion \citep{2012MNRAS.420..216S} 
provide modest fields estimates
$\sim 10^{13}$~G \citep{2012NewA...17..594C, 2012MNRAS.421L.127P, 2014EPJWC..6402002P}.  
The latter results are in better correspondence with the expected evolution of magnetic fields of NSs, as in modern scenarios initially large fields rapidly decay down to the level typical for normal radio pulsars \citep{2013NatPh...9..431P, 2013MNRAS.434..123V}.     

If an accreting NS is a member of a low-mass X-ray binary system (LMXBs),
then its age can be very large --- up to billions of years. It is hard to
imagine that a NS can still have strong magnetic field at such age. 
On the other hand when NS has a massive companion (high-mass X-ray binary --- HMXB) its
age is usually restricted to a few tens of Myrs which is still a large value in comparison with ages of most known isolated magnetars \citep{2015RPPh...78k6901T}.

The X-ray pulsar 4U0114+65 is one of the slowest known HMXRBs \citep{1996A&A...311..879R}. In the recent article by \cite{2017arXiv170604907S}, the long spin period (9.4 ks) and small emitting area of this sources were explained due to a magnetar-type magnetic field even in the frame of the wind settling accretion. The source is at a significant distance from any star formation region and the Galactic plane which suggests its large kinematic age. The concept of the kinematic age is precious for the studies of the neutron star properties because it measures the time since the supernova explosion which imparts the kick velocity to the system. 
The  estimate of the kinematic age is obtained by backward orbit integration. Such estimate should take into account the uncertainties in the proper motion measurements and unknown birth position. To deal with these we develop the Bayesian approach which allows us to quantify both uncertainties. 

Accreting magnetars have been also proposed  to explain properties of ultra-luminous X-ray sources (ULXs, see a review in \citealt{2017arXiv170310728K}) with NSs. 
The first of such source has been found by \cite{2014Natur.514..202B}, later two other examples were discovered by  
 \cite{2017MNRAS.466L..48I,2017Sci...355..817I}. To explain both timing and luminosity of such sources a large dipolar magnetic field is sometimes required, for example to support the accretion column which allows higher luminosity  (see \citealt{2015MNRAS.454.2539M} for such scenario). 

Known NS-ULXs belong to the class of HMXBs with Roche lobe overflow \citep{2014Natur.514..198M}, so it can be expected that compact objects in these systems have ages at least about several Myrs \citep{2017arXiv170310728K, 2008ApJ...675.1067F, 2011ApJ...734...23G}.
In this note we discuss parameters of NSs with which it is possible to obtain accreting magnetars in HMXBs in the framework generally consistent with rapid field decay in young magnetars such as soft gamma-ray repeaters and anomalous X-ray pulsars.

According to many calculations (see
e.g. \citealt{2015SSRv..191..315M} and references therein) 
at ages around few Myr the initially strong ($\sim 10^{15}$~G) dipole magnetic field decays by
several orders of magnitude from its initial value. 
In order to preserve a field $\sim 10^{14}$~G up to ages $\gtrsim$~few Myrs NS should satisfy a few conditions regarding the magnetic field evolution. These are properties of the Hall cascade in the NS crust and material impurity.

The article is structured as follows. In the second Section we introduce our formalism to describe the magnetic field evolution in a NS and identify the crucial terms responsible for the field evolution during first 10 Myr.
Results of the field evolution calculations are presented in Section 3. In the 4th Section we introduce the Bayesian estimate for the kinematic age and demonstrate that the NS in the accreting magnetar candidate 4U 0114+65 is at least 2 Myr old.
In Section 5 we discuss some additional topics related to our study. Finally, in the 6th Section we summarize our results. A list of accreting magnetar candidates and the formalism to estimate magnetic field from spin parameters are given in Appendix A and B, correspondingly. 

\section{Model of magnetic field evolution}
\label{sec:field}

The instantaneous magnetic field $B(t)$ depends on the initial value $B_0$ and follows a complicated evolution. To describe it theoretically we start with the formula introduced by \cite*{2008ApJ...673L.167A}:
\begin{equation}
B(t) = \frac{ B_0 \times
\exp(-t/\tau_\mathrm{ohm})}{1 + (\tau_\mathrm{Ohm}/\tau_\mathrm{Hall})
[1 - \exp(-t/\tau_\mathrm{Ohm})]}.
\label{magnet_decay}
\end{equation}

In this equation two distinct time scales are defined. The first one is related
to  the Ohmic decay (resistivity in the crust), $\tau_\mathrm{Ohm}$, and the second one --- to the Hall     
cascade, $\tau_\mathrm{Hall}$. The Hall evolution is in principle non-dissipative, however it redistributes the magnetic energy from high spatial scale (dipole field) to small scales (multipoles of higher order) which causes the decay of the dipole component and enhances release of magnetic energy.
The eq. (\ref{magnet_decay}) can be modified to include
some minimal value of the field, at which the decay is saturated which is usually attributed to the influence of the core magnetic field. As we are not interested in a long-term evolution ($\gtrsim 10^8$~yrs), we do not discuss this topic further, and omit possible saturation field. Note, that both time scale,  $\tau_\mathrm{Ohm}$ and $\tau_\mathrm{Hall}$, evolve with time and the latter one depends on the magnetic field value itself. Below we write equations for both time scales and choose parameters in such a way to reproduce recent detailed simulations of magneto-thermal evolution in the crust.

The timescale of the Hall evolution is:
\begin{equation}
\tau_\mathrm{Hall} = \frac{4\pi e n_eL^2}{cB(t)},
\end{equation}
with $n_e$ is local electron density, $e$ is the elementary charge, $B$ is
local instantaneous magnetic
field, $L$ is the typical spatial
scale of electric currents (it can be, for example, the local pressure height scale, see
\citealt{2004ApJ...609..999C}),
and $c$ is the speed of light.  
We can also define the Hall timescale using its initial value and the instantaneous magnetic field:
\begin{equation}
\tau_\mathrm{Hall} = \tau_\mathrm{Hall, 0} \frac{B_0}{B(t)}.
\end{equation}
Here it is assumed that $n_e$ and $L$ are constant. 

The Hall cascade can be terminated if so-called {\it Hall attractor} stage is reached. This stage was proposed by \cite{2014PhRvL.112q1101G,2014MNRAS.438.1618G} and then confirmed by \cite{2015PhRvL.114s1101W}. \cite{2014PhRvL.112q1101G} demonstrated that the stage is reached after a few initial Hall time scales. For a NS with initial field $\sim10^{14}$~G it happens after approximately a few hundred thousand years (up to 1 Myr). In our model we assume that the Hall attractor stage starts after three initial Hall time scales are passed.
As soon as the attractor is reached $\tau_\mathrm{Hall}$ is set to infinity, and the following field evolution proceeds only via Ohmic processes.

The Ohmic decay proceeds on two timescales $\tau_\mathrm{Ohm, ph}$ due to electron scattering on phonons, 
and $\tau_\mathrm{Ohm, Q}$ due to resistivity caused by the crust impurity.
The general form to describe the timescale of the Ohmic decay is:
\begin{equation}
\tau_\mathrm{Ohm} = \frac{4\pi \sigma L^2}{c^2},
\end{equation}
where $\sigma$ is the local electric conductivity which depends on resistivity agent.

The local electric conductivity is computed as:
\begin{equation}
\sigma = \frac{\sigma_\mathrm{Q} \sigma_\mathrm{ph}}{\sigma_\mathrm{Q} +\sigma_\mathrm{ph}}.
\end{equation}
Thus, for the timescales we can write $\tau_\mathrm{Ohm}^{-1}=\tau_\mathrm{Ohm, ph}^{-1}+\tau_\mathrm{Ohm, Q}^{-1}$.

The conductivity due to impurities is described as:
\begin{equation}
\sigma_{Q} = 4.4\times 10^{25} {\mathrm s}^{-1} \left( \frac{\rho_{14}^{1/3}}{Q}\right) \left(\frac{Y_e}{0.05}\right)^{1/3}\left( \frac{Z}{30}\right), 
\end{equation} 
according to \cite{2004ApJ...609..999C}.  In this equation  $\rho_{14}$ is the density in units $10^{14}$~g~cm$^{-3}$, and $Y_e$ is the electron fraction in the current layer.
The parameter $Q$ characterizes how ordered the crystalline structure of the crust: $Q=n_\mathrm{ion}^{-1}\Sigma_i\,n_i \times(Z^2-\langle Z\rangle^2)$. Here $Z$ is ion charge, and $n$ number density. 

A larger value of $Q \gg 1$ means that the crust composition is strongly non-homogeneous. The electrons are  scattered much more often in this case which significantly reduces the conductivity. For parameters of interest we obtain $\tau_\mathrm{Ohm, Q}=2\times 10^6 \, \mathrm{ yrs} \, Q^{-1}$, and we use this estimate below for different values of  $Q$. 

The phonon conductivity is computed as:
\begin{equation}
\sigma_\mathrm{ph} = 1.8\times 10^{25} {\mathrm s}^{-1} \left(\frac{\rho_{14}^{7/6}}{T^2_8}\right)\left(\frac{Y_e}{0.05}\right)^{5/3},
\end{equation}
The value $T_8$ is the temperature of the crust in units $10^8$~K. Our choice of  parameters is guided by detailed numerical simulations by \cite{2013NatPh...9..431P}. For magnetars the layer in the crust which controls the long-term field evolution is $\rho_{14}=0.8$. The electron fraction seems to be a factor of 2 larger in \cite{2013NatPh...9..431P} comparatively to \cite{2004ApJ...609..999C}. The phonon conductivity goes to infinity when the temperature of the crust drops below $T_\mathrm{U}$. In our calculations we use $T_\mathrm{U}=2.6\times 10^7$~K. 

To calculate $\tau_\mathrm{Ohm, ph}$ we need to know the temperature in the crust. 
For NSs at the stage of Hall cascade we use the following analytical fit for
the crustal temperature calculated by \cite{2013MNRAS.434..123V}:

\begin{equation}
T=T_1\, \exp(-t/\tau_1) +T_2\, \exp(-t/\tau_2). 
\label{t1}
\end{equation}
Parameters $T_1, T_2, \tau_1, \tau_2$ depend on the initial magnetic field and on the NS mass (massive NSs in which direct URCA processes are allowed, cool faster). 
For $B_0=10^{15}$~G we take $T_1=7\times 10^8$~K,~$\tau_1=150$~yrs,
$T_2=1.5\times 10^8$~K, $\tau_2=2.5\times 10^6$~yrs. For smaller fields $T_2$
and $\tau_2$ are smaller (i.e., cooling proceeds more rapidly due to smaller energy release due to field decay). For initial fields $\lesssim$~few~$ 10^{13}$~G
additional heating is not important. As soon as the temperature is determined we calculate the timescale via $\tau_\mathrm{Ohm, ph}=2\times 10^6 \mathrm{ yrs} \, T_8^{-2}$.
Magnetars are known 
sources of thermal X-ray emission which is explained by their high surface temperature. The exact mechanism causing this heating is unknown 
\citep{2016ApJ...833..261B}. One of possible alternatives is the heating produced by crustal electric current \cite{2013MNRAS.434..123V}
which is especially efficient during the Hall cascade. 

When the Hall attractor stage is reached rapid dissipation of the magnetic
field energy is over, and the crust quickly relaxes to the stage without
additional heating. In this case we use an analytical approximation for
cooling tracks from \cite{2011MNRAS.412L.108S}:

\begin{equation}
T=b \left( \frac{t}{1 \mathrm{yr}} \right)^a \exp(-t/\tau_\mathrm{c}).
\label{e:std_cooling}
\end{equation}  
Parameters are chosen to be: $b=6.56\times10^8$~K, $a=-0.185$, and
$\tau_\mathrm{c}=8.58\times 10^5$~yrs. This fits a NS without direct URCA processes in the core.

In eq. (\ref{magnet_decay}) the instantaneous magnetic field is used in the left and in the right hand side. To express it explicitly we need to solve a quadratic equation:
$$
B^2(t) \left(\frac{\tau_\mathrm{ohm}}{\tau_\mathrm{Hall, 0} B_0}\right) \left[1 - \exp\left(-\frac{t}{\tau_\mathrm{ohm}}\right)\right] +B(t)\hspace{1.5cm}
$$\begin{equation}
\hspace{4cm}- B_0 \exp\left(-\frac{t}{\tau_\mathrm{ohm}}\right)=0
\end{equation}
The solution is:
\begin{equation}
B(t) = \frac{B_0}{2} \left( -\frac{1}{\gamma(t)} + \sqrt{\frac{1}{\gamma^2(t)} + \frac{4\kappa(t)}{\gamma(t)}} \right); 
\label{e:solution}
\end{equation}written by means of the auxiliary variables:
\begin{equation}
\kappa(t) = \exp\left(-\frac{t}{\tau_\mathrm{Ohm}}\right), 
\end{equation}and
\begin{equation}
\gamma(t) = \left(\frac{\tau_\mathrm{Ohm}}{\tau_\mathrm{Hall, 0}}\right) \left[1-\kappa(t)\right].
\end{equation}
The exact algorithm that we use is as follows: first, we check if the Hall attractor is reached i.e. $t > 3\tau_\mathrm{Hall, 0}$. If it is the case we set $\tau_\mathrm{Hall}=\infty$, otherwise we compute the actual Hall timescale. To avoid an unphysical jump in $B(t)$ at the moment when the Hall attractor starts operating, we substitute new $B_0$ in eq. (\ref{e:solution}) which is equal to the last moment before the turn-off of the attractor.
Second, we compute the temperature according to eq. (\ref{t1}) or eq.(\ref{e:std_cooling}) depending on whether the Hall attractor is reached. If the temperature is
larger than $T_U$, we compute $\tau_\mathrm{ohm,ph}$. If the $T < T_U$ we set $\tau_\mathrm{ohm,ph} = \infty$.  Third we substitute all timescales into eq.(\ref{e:solution}). 
The instantaneous magnetic field $B(t)$ is computed then at a time grid. 


\section{Results of magnetic field evolution calculations}
\label{sec:results}

We made runs for different sets of parameters determining the magnetic field evolution. Here we  present results
for our reference model, in which 
$\tau_\mathrm{Hall}=10^4 \, \mathrm{yrs} \, (10^{15}\, {\mathrm G}/B)$ and 
$\tau_\mathrm{Ohm,ph}=2\times10^6\, \mathrm{yrs} \, T_8^{-2}$. 
In the models with the Hall attractor, we turn it on at $t=3\tau_\mathrm{Hall, 0} = 3\times 10^4 \, \mathrm{yrs} \, (10^{15}\, {\mathrm G}/B_0) $. We perform our simulations for three values of $Q$: 1, 10, and 100 and it is kept constant.

In Fig. ~1 we present results for  the initial field $B_0=10^{15}$~G. These refer to  a NS without direct URCA processes in the core.

In addition to three tracks for different $Q$, in Fig.~1 we also plot a curve for the case without the Hall attractor (upward triangles). This line  is calculated with $Q=1$. However, without termination of the Hall cascade even for low $Q$ it is impossible to save large field at ages $\gtrsim 1$~Myr. In this case thermal evolution always proceeds according to eq. (\ref{t1}), i.e. scattering on phonons is active up to several Myrs (temperature is above $T_\mathrm{U}$).

\begin{figure}
\center{\includegraphics[width=84mm]{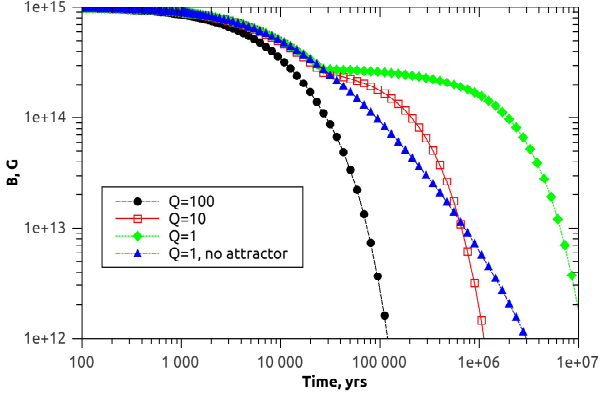}}
\caption{Magnetic field decay for several sets of parameters. Initial field $B_0=10^{15}$~G.
Filled circles, empty squares, and filled diamonds corresponds to 
the standard case with Hall attractor (since $t=3\tau_\mathrm{Hall, 0}$)
and different values of $Q$ (see the legend). Upward filled triangles
correspond to the model with $Q=1$ and no Hall attractor. In the latter case
thermal evolution always proceeds along the track fitted by the sum of two
exponents (see text).
}
\label{bevol}
\end{figure}

\begin{figure}
\center{\includegraphics[width=84mm]{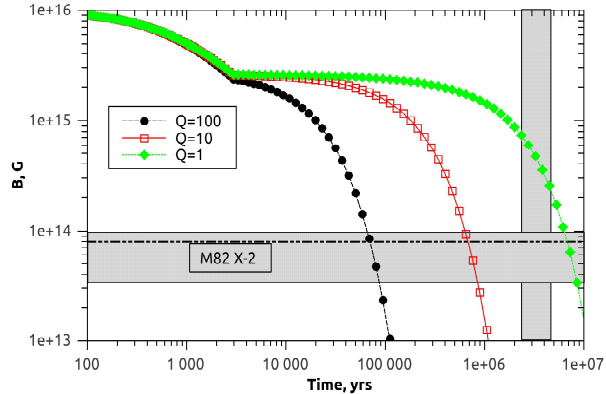}}
\caption{Magnetic field decay for several sets of parameters and estimates for known sources. Initial field $B_0=10^{16}$~G.
Filled circles, empty squares, and filled diamonds corresponds to 
the standard case with Hall attractor (since $t=3\tau_\mathrm{Hall, 0}$)
and different values of $Q$ (see the legend). Dot-dot-dashed horizontal line corresponds to the magnetic field estimate for the ULX M82 X-2 from Mushtukov et al. (2015): $B=8\times 10^{13}$~G. Horizontal grey box corresponds to the field estimate for the NS in 4U 0114+65 by Sanjurjo et al. (2017): $B\sim (3-10)\times10^{13}$~G. Vertical grey box shows credential interval which contains $95\%$ probability for the age of the NS in this source: 2.4-5 Myrs (this work).
}
\label{bevol16}
\end{figure}

We expect that NSs in accreting magnetar systems are at least older than 1
Myr. Thus, as it is visible from Fig.~1, in most of the cases the remaining
magnetic field is $\lesssim 10^{13}$~G. However, we can construct an
evolutionary track for the field which allows values $\sim 10^{14}$~G
several Myrs after the NS formation. Two main ingredients are: the Hall attractor
and low $Q$. In addition, it is necessary that resistivity due to phonons is low (i.e., the crust is colder than $T_\mathrm{U}$) during most of the evolution (say, after few hundred thousand years). 

This combination of parameters is not the expected one, as typically it is
assumed (\citealt{2013NatPh...9..431P}) that $Q$ is large for magnetars, because currents are situated in deep crustal layers in the zone of nuclear pasta, where impurities are important.    

For comparison in Fig.~2 we present magnetic field evolution for $B_0=10^{16}$~G. For such large fields results are not very sensitive to the choice of coefficients in eq. (8). This is so because for higher field the Hall attractor stage starts very early, and also at early phases of evolution decay is mostly driven by the Hall term. With respect to Fig.~1 curves are shifted  not only up, but also to the left, as the initial evolution proceeds much faster for larger fields due to smaller Hall time scale. Later evolution, at ages $\gtrsim$~100 kyrs, is not much changed. Obviously, it is still impossible to explain accreting magnetars without involving the Hall attractor and low values of $Q$ even for very large initial magnetic fields. 

To make good estimates of $Q$ or at least to put strict limits on its value, it is necessary to use sources with known ages of NSs. In many cases just very approximate estimates are available from analysis of binary evolution. However, in a few cases it is possible to derive age estimates from kinematics of well-studied binaries in the Galaxy. In the following section we provide such calculations for the X-ray binary 4U 0114+65.

\section{Age of 4U0114+65}
The accreting magnetar candidate 4U 0114+65 is an excellent source to place some limits on the inner crust impurity. The magnetic field of this source was recently estimates as $\sim 10^{14}$ G (see Introduction). Moreover, the source  is at substantial offset from any star-forming region which is most probably caused by a velocity kick imparted to the system during the first supernova explosion.  The large OB association CAS OB8 \citep{1982BICDS..22..132R,1970csca.book.....A} is $2^\circ$ away which is comparable with the distance of  4U0114+65 from the Galactic plane ($b = +2^\circ .5635$).  At angular separation of one degree an old stellar cluster Pfleiderer 1 with the age 1 Gyr can be found \citep{2012A&A...543A.156K, 2013A&A...558A..53K,2014A&A...568A..51S,2014yCat..35680051S}.  Clusters of such age are not associated  with OB stars.
The source is at $\alpha'$ = $01^h$ $18^m$ $02^s.6974$  $\delta'$=$+65^\circ$ $17'$ $29.''830$\footnote{The measured quantities are written with prime here}  and has effectively an upper limit on parallax $\varpi' = 0.11$ mas set by {\it Gaia} with its accuracy 
0.23 mas in the first data release \citep{2016A&A...595A...4L}. The parallax indicates a distance larger than 4 kpc  which is in
agreement with photometric distance $7\pm 3.6$ \citep{1996A&A...311..879R} based on 
the apparent magnitude $m_v = 11.14$, $E(B-V)=1.24$ and spectral type of the companion B1Ia.  According to the recent three-dimensional map of the Milky Way dust \citep{2015ApJ...810...25G} the measured reddening of  $E(B-V)=1.24$ corresponds to distances in range 3.8-6.0 kpc in the direction to 4U0114+65.

The system 4U0114+65 has measured proper motion $\mu_{\alpha *}' = -1.4\pm 1.72$ mas/year, $\mu_\delta' = 3.17\pm 1.56$ mas/year \citep{2007A&A...474..653V} and the radial velocity $v_r' = -57\pm 2$ km/s (\citealt{2004A&A...424..727P, 1985ApJ...299..839C}, the observational uncertainty is made larger to take both fits into account). Such proper motion in combination with the angular separation of $2^\circ$ easily gives the kinematic age of order 2 Myr irrespectively of the source distance.

\begin{figure}
\center{\includegraphics[width=84mm]{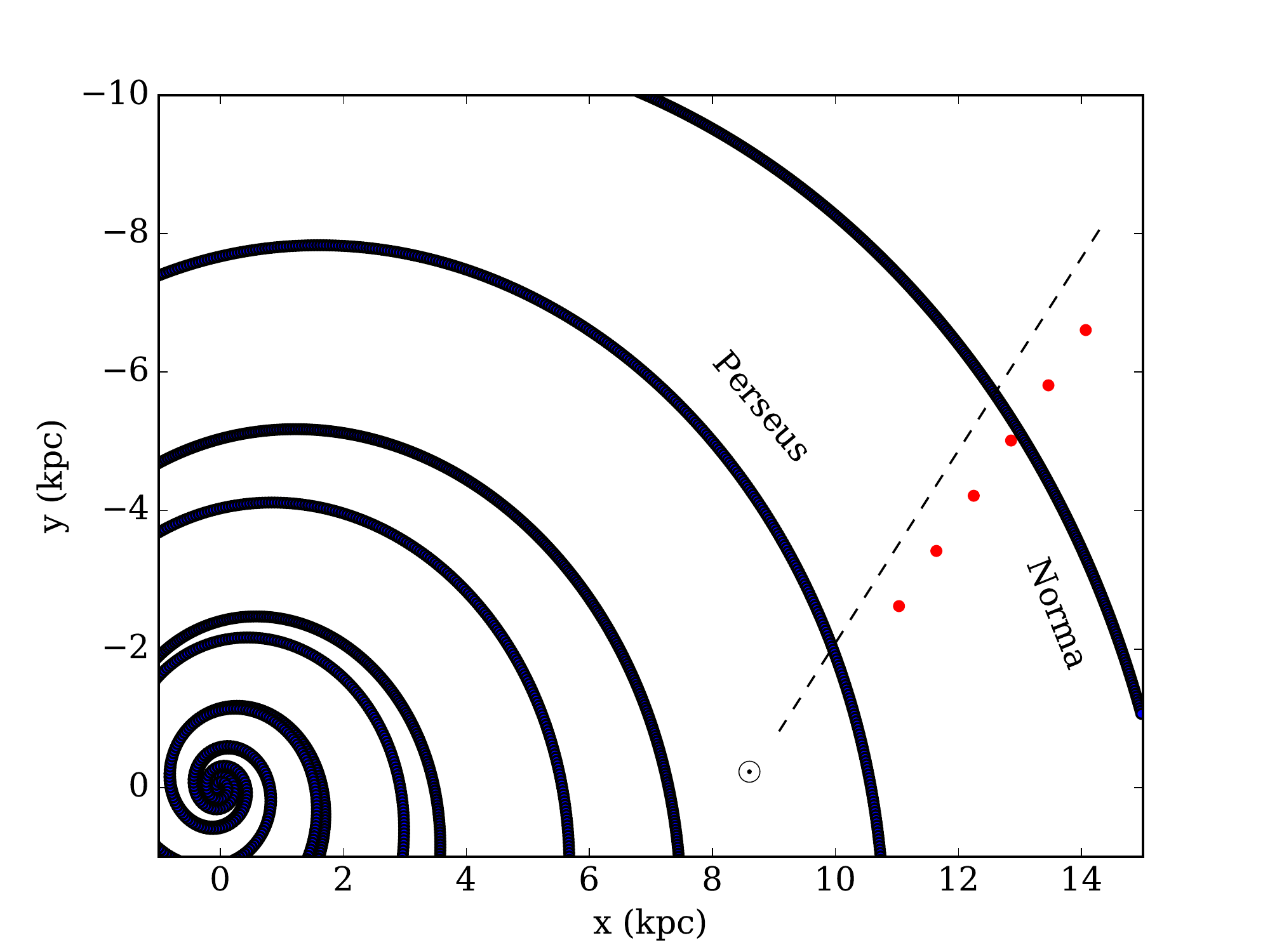}}
\caption{The Galactic spiral pattern and the direction toward 4U0114+65 (dashed line). Dots show possible birth positions of the source obtained by back integration in time of its current location, proper motion and radial velocity for different assumed distances D in the range from 4 until 10 kpc from the observer with step 1 kpc. 
}
\label{f:spiral}
\end{figure}

To better understand the possible origin of the source and its kinematic age we plot the direction to 4U0114+65 on top of the four spiral arms based on \cite{1992ApJS...83..111W} and \cite{1976A&A....49...57G}, see Fig. \ref{f:spiral}. 
Given its distance, the system is most probably originated in the Norma spiral arm. It allows us to get some estimate of the system age. Two approaches are described in the following sections: (1) the classical backward orbit integration for a number of distances; (2) the Bayesian approach. 

\subsection{Backward orbit integration}
The kinematic age of a system is usually considered as a time which is required for the 
system to travel from its birth position (often assumed as the Galactic plane $z=0$) until its recent location, see e.g. \cite{2013MNRAS.430.2281N}. This approach is justified because majority of B stars are born in a very thin layer close to the Galactic plane (scale height 45 pc according to \citealt{2000AJ....120..314R}). It is quite common that some estimate of the object distance is available, but the radial velocity is missing. In our case the distance is unknown, but the radial velocity is perfectly constrained. This is the reason to start the backward orbit integration from different positions along the line of sight from 4 kpc with 1 kpc spacing between separate initial conditions. 

The equatorial coordinates, distance, proper motion and the radial velocity uniquely constrain the initial conditions for the orbit integration. The integration is performed by means of the {\sc galpy } python package \citep{2015ApJS..216...29B}\footnote{ http://github.com/jobovy/galpy} using the second gravitational potential from the paper by \cite{2013A&A...549A.137I}.
The conversion to  units\footnote{Units in the Galactic dynamics are determined by the choice of the factor for the total gravitational potential to assure that $v_\mathrm{circ}(r)=r = 1$.} used in the {\sc galpy} is done assuming the Solar distance $R_\odot = 8.5$ kpc and the Solar velocity $v_\odot = 220$ km/s. 
The orbit is integrated backward in time from the current position of the binary until a moment when it crosses  the Galactic plane ($|z |< 10$~pc which gives the most probable age), an age when it approaches the height $z = 100$ pc is also reported (two times of the scale height from \citealt{2000AJ....120..314R} which gives a lower limit on the age). The orbit intersections with the Galactic plane for different distances are shown with dots in Fig.~\ref{f:spiral}. The Norma spiral arm seems to be the most prominent formation region (see also \citealt{1996A&A...311..879R}).  

For the whole range of distances the age estimate is exactly the same and it is equal to 3.48 Myr. It happens because the intersection time is defined by relation between $z$ and $v_z$ and the contribution of
the radial velocity is constant which makes both $z$ and $v_z$ linearly depending on distance. The intersection with $z=100$ pc leads to a lower limit for the age ranging from 1.97 Myr for smaller distances up to 2.72 Myr for larger distances. The age depends on distance in this case because characteristic height $z=100$ pc has different angular size at different distances from us. 

The distance from the Galactic center $R_0=13.8$ kpc and azimuth $\phi=21^\circ .30$ is close to the Norma spiral arm (the {\sc galpy} uses the left-handed frame for the orbit integration). The position and velocity found in the backward integration is used then in the next subsection as the first guess for the Markov Chain Monte Carlo process.  

\subsection{The Bayesian age estimate}
The proper motion of the system 4U0114+65 is measured with a significant uncertainty which leads to a large family of possible orbits. 
We introduce here the Bayesian approach for the kinematic age estimate. We start from the conditional probability to obtain measurements given the actual values for the birth position $\vec R_0 = [R_0, \phi_0, z_0]$ (radial distance $R_0$, azimuth $ \phi_0$ and height above the galactic plane $z_0$), and three dimensional velocity $\vec v_0 = [v_{r,0}, v_{T, 0}, v_{z,0}]$ (radial $v_{r,0}$, transversal $ v_{T, 0}$ and vertical $v_{z,0}$) as well as the system age $t$ since the moment of the first supernova explosion. The conditional probability is:
$$
p(\mu_{\alpha *}', \mu_\delta', v_r', \alpha', \delta' | \vec R_0, \vec v_0, t ) \propto 
$$
\begin{equation}
g(\mu_{\alpha *}' | \mu_{\alpha *})g(\mu_\delta'|\mu_\delta ) g(\alpha' | \alpha) g(\delta'|\delta) g(v_r' | v_r)
\label{e:likel}
\end{equation}
where $g(x'| x)$ is a Gaussian in the form:
\begin{equation}
g(x'| x) = \frac{1}{\sqrt{2\pi}\sigma_x}\exp\left(-\frac{(x' -x)^2} {2\sigma_x^2}\right).
\end{equation}
The values with prime are used to show measured quantities while values without prime are for actual (unknown) values. The difference between the measured and actual values appears only because of the observational errors.
Eq. (\ref{e:likel}) is essentially the likelihood which constrains the possible birth properties of the system such a way that its current sky position, proper motion and the radial velocity are in agreement with observations. In the case of the coordinates $\alpha,\delta$ the observational uncertainty is artificially increased up to $1'$ because more accurate precision is not necessary. 
The posterior can be written as multiplication of the likelihood to prior:
$$
P(\vec R_0, \vec v_0, t | \mu_{\alpha *}', \mu_\delta', v_r', \alpha', \delta') \propto
$$
\begin{equation}
p(\mu_{\alpha *}', \mu_\delta', v_r', \alpha', \delta' | \vec R_0, \vec v_0, t )  f(\vec R_0, \vec v_0, t ).
\end{equation}
The normalization is not important here because it is a constant. 
The prior $ f(\vec R_0, \vec v_0, t )$ is a multiplication of three independent priors: for the Galactic structure $f_G (\vec R_0)$ which includes the description of the spiral pattern, for the initial systemic velocity $f_v(\vec v_0)$ and a flat Jefferson prior for time in the range from 0.01 Myr till 100 Myr. 

The prior for the spiral pattern is:
$$
f_G(\vec R_0) \propto \frac{1}{\sqrt{2\pi} \sigma_\mathrm{r} \sigma_\mathrm{z}} \exp\left(-\frac{(R_0 - r_k \exp((\phi_0 - \phi_\mathrm{k})/\kappa_\mathrm{k}))}{2\sigma_\mathrm{r}^2}\right)
$$
\begin{equation}
\hspace{3.5cm}\times\exp(-z/\sigma_\mathrm{z})
\end{equation}
This complicated function represents the logarithmic spiral with $r_\mathrm{k} = 3.48$ kpc, $\kappa_\mathrm{k} = 4.25$ and $\phi_\mathrm{k} = 2\pi$ in the case of the outer part of the Norma arm \citep{1992ApJS...83..111W}. The typical dispersions are selected as $\sigma_\mathrm{r} = 0.35$ kpc \citep{2006ApJ...643..332F} and $\sigma_\mathrm{z} = 0.045$ kpc \citep{2000AJ....120..314R}.

The prior for the birth kick velocity is a simple isotropic Maxwellian with reduced $\sigma = 150$ km/s to take into account that binaries can be disrupted and the natal kick velocity of the neutron star is not the systemic velocity of the binary, see e.g. \cite{2017MNRAS.467..298R}.

Since we are interested only in the kinematic age, all spatial and dynamical variables are integrated out:
\begin{equation}
P(t) \propto \int ... \int P(\vec R_0, \vec v_0, t | \mu_{\alpha *}', \mu_\delta', v_\mathrm{r}', \alpha', \delta') d^3 \vec R_0 d^3 \vec v_0
\end{equation}
The simplest way to implement this multidimensional integration is to use the Markov chain Monte Carlo sampler. The simulations are performed for 48 ``walkers'' (independent Markov chains) and 4000 is the number of iterations with first 1000 iterations were excluded to allow the process to converge to the stationary distribution. For this process the maximum radial distance was set to 15 kpc and maximum velocity to 300 km/s in each direction. To quantify the posterior distribution the $95\%$ credential interval is computed for samples. Analytically this interval is described as:
\begin{equation}
\int_a^b P(t) dt = 0.95
\end{equation}
where $a$ and $b$ are the boundaries of the interval. The credential interval ranges from 2.39 Myr till 4.96 Myr. The posterior peaks at 3.90 Myr. A use of the velocity prior in form of Maxwellian with $\sigma = 250$ km/s (typical for isolated neutron stars) extends the credential interval by $\approx 0.5$ Myr at both sides: 1.92 Myr to 5.65 Myr with peak at 4.0 Myr.   

We conclude this section with the statement that taking into account the age estimates presented above, properties of 4U 0114+65 as an accreting magnetar candidate can be explained with initial fields $\sim10^{15}$~--~$10^{16}$~G and $Q\sim 1$~--~5, see Fig.~\ref{bevol16}.

\section{Discussion}
\label{sec:disc}


Accreting magnetars remain hypothetical sources, i.e. estimates of magnetic field of NSs in candidate systems are not certain. Still, several authors discussed the origin and evolution of such binaries. Above we focused on the magnetic field evolution to study under which conditions the field can remain high enough for required long time. 

For the first time we tried to model NS magnetic field evolution for accreting magnetars in the framework used for studies of standard isolated magnetars.
In our analysis we did not include possible influence of accretion on the field decay. If this effect is taken into account (see, for example, \citealt{2016MNRAS.461....2P}) then the field might be even lower than in our estimates. I.e., smaller values of $Q$ might be appropriated to fit properties of the systems discussed in this paper. However, NSs in HMXBs are relatively young, and if we are not dealing with ULXs, then the amount of accreted matter is not that high ($\lesssim$~few~$\times 0.001\, M_\odot$) to result in significant additional field decay.

In our calculations we assumed the the Hall attractor stage starts at $t=3\tau_\mathrm{Hall, 0}$. According to \cite{2014PhRvL.112q1101G} and \cite{2015PhRvL.114s1101W} the onset of this stage is not so certain. It can start later. In this case we present conservative estimates, i.e. for later Hall attractor initiation the magnetic field might decay more (see Fig.~1 for the case of without the Hall attractor). Thus, it is necessary to use even smaller $Q$ to explain accreting magnetars with ages from few Myrs up to few tens of Myrs.


\cite{2015ApJ...802..131S} studied possible evolutionary channels to explain ULXs with NSs. According to this study typical ages of NSs at the time when the Roche lobe overflow is initiated are about several tens of Myrs.  \cite{2015ApJ...802L...5F} came to similar conclusions. The system M82 X-2, according to these authors, is most probably $\lesssim$~65 Myrs old, and the NS progenitor had a mass $8$~--~$25\, M_\odot$; thus the NS might have an age  $\gtrsim$~few tens of Myrs. From Fig.~2 it is visible, that with $Q=1$ we can explain the field estimate for the source M82 X-2 made by \cite{2015MNRAS.454.2539M} just for age $\lesssim 10^7$~yrs even with $B_0=10^{16}$~G. For larger ages 
it is necessary to use lower values of $Q$, which can be applicable to normal radio pulsars, but is not considered to be typical for magnetars.
Better (and more numerous) estimates of ages of NSs in accreting magnetar candidate systems might help to improve our understanding of their magnetic fields evolution.

\section{Summary}\label{sec:summary}

Accreting magnetars have been proposed as a class by \cite{2012MNRAS.425..595R}, and later on ULXs with NSs were suggested an possible members of this group \citep{2015MNRAS.448L..40E}. Despite the fact that evidence in favour of their existence is up to date just indirect, such systems might be formed if a NS remains highly magnetized after tens of Myrs of evolution. It is possible to find a set of parameters which allows this.

To better constrain the NS parameters responsible for the field evolution it is necessary to have better estimates of ages for accreting magnetar candidates.
We suggest a new Bayesian estimate of the kinematic ages. Applying this estimate to the accreting magnetar candidate 4U0114+65 with realistic priors we find that its kinematic age is is 2.4-5.0 Myr ($95\%$ credential interval).

We conclude, that to form an accreting magnetar with an age $\gtrsim$~few Myrs it is necessary to include three main ingredients: the Hall attractor, absence of scattering on phonons after few hundred thousand years, and low ($\lesssim$~few) value of parameter $Q$ which characterizes the role of impurities.

\section*{Acknowledgements}
The authors are grateful to Daniele Vigano, Jose Pons, and Peter Shternin for the opportunity to use their calculations of thermal evolution of neutron stars. We thank prof. Frank Verbunt and anonymous referee for their comments on the manuscript. S.B.P. also thanks Peter Shternin for discussions.
The work of  A.P.I. is supported by NOVA PhD funding.
S.B.P. acknowledges support from RSF grant No. 14-12-00146.
This research has made use of NASA's Astrophysics Data System.




\bibliographystyle{mnras}
\bibliography{magn_bin}


\appendix 

\section{Proposed accreting magnetars}

Below we list some of proposed accreting magnetar candidates:

\begin{itemize}
\item ULX. NuSTAR J095551+6940.8 (M82 X-2). \cite{2015MNRAS.448L..40E}.
\item ULX. NGC 5907. \cite{2017Sci...355..817I} 
\item ULX. NGC 7793 P13. \cite{2017MNRAS.466L..48I}.
\item 4U0114+65. \cite{2017arXiv170604907S}.
\item 4U 2206+54. \cite{2010Ap.....53..237I}.
\item SXP1062. \cite{2012ApJ...757..171F}
\item Swift J045106.8-694803. \cite{2013MNRAS.428.3607K}.
\end{itemize}

Also a large list of possible candidates can be found in \cite{2014MNRAS.437.3863K, 2014MNRAS.437.3664H} (see also \citealt{2015ApJ...813...91S}). These candidates are selected on the base of timing properties of X-ray pulsars.

Individual estimates of magnetic field can be very different for a given source as several models (and considerations) can be applied. For example, in the case of M82 X-2, which is the most famous source in the list, estimates range from standard fields $\sim 10^{12}$~G \citep{2016arXiv160607096C} up to $\sim 10^{14}$~G \citep{2016MNRAS.457.1101T}, including the case of normal dipole ($\sim 10^{12}$~G) but strong multipole fields ($\sim 10^{14}$~G) \citep{2015AN....336..835T}.

For several other sources (for example, IGR J16358-4726 and 4U 1954+319, see \citealt{2014ApJ...786..127E}) it was noted that basing on the model of standard disc accretion \citep{1979ApJ...232..259G} NSs in these systems might have magnetar-scale dipolar fields. However, more detailed analysis usually demonstrate that sources can be explained with a different model of accretion, in which there is no necessity of strong magnetic field. 
Thus, independent measurement of magnetic fields in such sources is of interest for accretion physics. 

\section{Magnetic field estimates based on timing properties}

 Here we briefly remind basics of magnetic field estimates from data on spin period, $p$, and period derivative, $\dot p$. We basically follow \cite{2012NewA...17..594C}.
 
 Magnetic field can be estimated either under so-called hypothesis of equilibrium period, or from period variations (spin-up or spin-down) for a specified model of accretion. 

Assuming that the spin period of a NS is equal to its
equilibrium period, the magnetic field $B$
for disc accretion can be estimated as follows:
\begin{equation}
B=2^{-1/4} \pi^{-7/6} k_\mathrm{t}^{-7/12} \epsilon^{7/24} p^{7/6}{\dot M}^{1/2}(GM)^{5/6} R^{-3}.
\end{equation}  
Here $\dot M$ is the accretion rate, $M$ and $R$ are the NS mass and radius, and $k_\mathrm{t}$ and $\epsilon$ are coefficients of order unity (often used values are $\epsilon = 0.45$, $k_t=1/3$).

For wind accretion:
\begin{equation}
B=2\sqrt{\frac{2\eta}{k_\mathrm{t} \pi}} p^{-1/2}_\mathrm{orb}v^{-2}(GM)^{3/2}{\dot M}^{1/2} p R^{-3}.
\end{equation} 
Here $p_\mathrm{orb}$ is the orbital period of a binary, $v$ is the stellar wind velocity, and $\eta$ is a coefficient of order unity (often it is assumed  $\eta = 1/4$). 

In the model of settling accretion from stellar wind developed recently by \cite{2012MNRAS.420..216S} a different equation is valid:

\begin{equation} \label{E:shakura}
B=0.24\times 10^{12}\, {\mathrm G}\, \eta_\mathrm{s} \left(\frac{p/100 \rm{s}}{p_\mathrm{orb}/10 \rm{days}}\right)^{11/12}{\dot M_{16}}^{1/3}\left(v/(10^8\, \mathrm{cm/s})\right)^{-11/3},
\end{equation}
where $\eta_\mathrm{s}$ is a coefficient of order unity. This model is valid for relatively low luminosities, and it was successfully applied to many systems (see, for example, \citealt{2014EPJWC..6402002P, 2017arXiv170604907S} and references therein).

A NS star can be out of spin-equilibrium if it is rapidly spinning up or down. In this case it is possible to neglect either spin-up or spin-down torque. This allows us to estimate the magnetic field. For example, for disc accretion using the observed values of the maximum spin-up rate, the magnetic field of NS can be estimated as follows:

\begin{equation}
B=\frac{2^{4}\pi^{7/2}}{\epsilon^{7/4}} \frac{(I\dot p)^{7/2}}{R^{3}p^{7}\dot M^{3}(GM)^{3/2}},
\end{equation}
where $I$ is the moment of inertia of a NS. 

In the case of the maximum spin-down rate 
the magnetic field of a NS can be estimated as follows:
\begin{equation}
B=\frac{2}{R^3}\left(\frac{I \dot p GM}{2\pi k_\mathrm{t}}\right)^{1/2}.
\end{equation}
This estimate should be normally considered as a lower limit, 
since we cannot be sure that no accelerating torque
exists at that moment.

Note, that there are many more equations to estimate magnetic field under the hypothesis of spin equilibrium or without it. Description of some of them can be found in \cite{2015ApJ...813...91S}.

\end{document}